\documentclass[conference]{IEEEtran}
\usepackage{cite}
\usepackage{amsmath,amssymb,amsfonts}
\usepackage{algorithmic}
\usepackage{graphicx}
\usepackage{hyperref}
\usepackage{textcomp}
\usepackage{xcolor}
\usepackage{url}
\def\BibTeX{{\rm B\kern-.05em{\sc i\kern-.025em b}\kern-.08em
    T\kern-.1667em\lower.7ex\hbox{E}\kern-.125emX}}
\begin{document}

\title{Automatic Scam-Baiting Using ChatGPT}

\author{\IEEEauthorblockN{Piyush Bajaj}
\IEEEauthorblockN{Matthew Edwards} \\
\IEEEauthorblockA{\textit{School of Computer Science} \\
\textit{University of Bristol}\\
Bristol, UK \\
piyushbajaj71@gmail.com, matthew.john.edwards@bristol.ac.uk}}

\maketitle

\begin{abstract} 

Automatic scam-baiting is an online fraud countermeasure that involves automated systems responding to online fraudsters in order to waste their time and deplete their resources, diverting attackers away from real potential victims.  Previous work has demonstrated that text generation systems are capable of engaging with attackers as automatic scam-baiters, but the fluency and coherence of generated text may be a limit to the effectiveness of such systems. 

In this paper, we report on the results of a month-long experiment comparing the effectiveness of two ChatGPT-based automatic scam-baiters to a control measure. Within our results, with engagement from over 250 real email fraudsters, we find that ChatGPT-based scam-baiters show a marked increase in scammer response rate and conversation length relative to the control measure, outperforming previous approaches. We discuss the implications of these results and practical considerations for wider deployment of automatic scam-baiting.

\end{abstract}

\begin{IEEEkeywords}
fraud, scam-baiting, active defence
\end{IEEEkeywords}

\section{Introduction}
Email-based fraud is a major component of online crime. Investment scams, online
dating fraud, tech support scams, employment scams, lottery and inheritance
scams, and advanced fee fraud schemes are all commonly initiated through an
approach via email, and the latest IC3 report suggests up to 113,700 victims
have reported significant financial losses from these categories in
2022~\cite{computer_fraud_amp_security_2022}. Conviction rates for these
offences are notoriously low due to the transnational nature of offending, which
poses significant hurdles for prosecution. Traditional countermeasures have
focused on blacklisting originators of fraud~\cite{toorn} and building email filters that
prevent email users from being exposed to fraudulent approaches~\cite{malik}.

More recently, researchers have proposed paying greater attention to an
\emph{active defence} posture when combatting cybercrime~\cite{canham}. As an
implementation of this within the domain of email-based offending in particular,
Chen et al.~\cite{wentaopaper} have demonstrated the feasibility of
\emph{automatic scam-baiting}, in which an automated responder system replies to
emails from fraudsters in order to waste their time, distracting offenders from
real victims. However, the GPT-Neo text generation systems tested by Chen et al.
showed significant limits in their ability to generate coherent and persuasive
email messages, with conversations commonly ending due to the generation of a
poor-quality text sample~\cite{wentaopaper}.

With more powerful text generation systems now widely available, this paper
explores approaches to improving the art of automated scam-baiting. In
particular, we examine (a) whether and to what degree an updated text generation
system (ChatGPT) is more effective at initiating and elongating email
conversations with scammers and (b) whether a text generation system given
examples of human scam-baiting conversation to imitate will out-perform a text
generation system given only general scam-baiting instructions. We test our
systems in a month-long experiment involving randomised allocation of scammer
approaches to two ChatGPT-based reply systems and one template-based control
measure.  Alongside our quantitative results, we report on our observations from
the various scam-baiting exchanges, including both limitations and unexpected
benefits of the approach as well as potential tactical responses from offender
populations.

The rest of this paper proceeds as follows. In Section~\ref{sec:background} we
provide background on scam-baiting activities and outline why ChatGPT is a
promising candidate for application in this domain.  Section~\ref{sec:method}
details the approaches tested and describes our experimental deployment.
Section~\ref{sec:results} presents our main findings about the effectiveness of
the reply strategies, while Section~\ref{sec:quality} discusses additional
observations, limitations and potential future developments. We conclude with
our main recommendations for ongoing work in automatic scam-baiting.

\section{Background}
\label{sec:background} 

\subsection{Scam-baiting}

Scam-baiters are online volunteers who reply to fraudsters in the guise of
victims, in order to waste the fraudsters' time. While there can be a range of
personal motivations for scam-baiting~\cite{zingerle,twitch}, some of which have
been considered critically~\cite{beatings,zingerle,sorell}, most modern
scam-baiters justify their work on the grounds that the time and energy
fraudsters spend interacting with them is time not spent defrauding a
real victim.  Herley~\cite{herley} argues that by decreasing the density of viable
targets, scam-baiting activity can have a disproportionate impact on fraudsters 
reducing the cost effectiveness of their work. Observation also
suggests that scammers find dealing with scam-baiters
frustrating~\cite{matthew}, which could be expected to lead to demoralisation.

In some settings, scam-baiting has been employed as research tool to understand
elements of online fraud
offending~\cite{zbinden2023scambaiting,clayton,gallagher}, making this technique
an active extension of the classic honeypot investigative tool~\cite{honeypot}.
However, a key aspect of scam-baiting is that it forms a template for a form of
\emph{social engineering active defence}~\cite{canham}, in which social
engineering techniques are deployed against internet fraudsters in order to
counteract their offending.  Chen et al.~\cite{wentaopaper} suggest that
\emph{automatic} scam-baiting could be an effective tool for combatting online
fraud, highlighting the possibility of avoiding the human costs of manual
scam-baiting. Their work demonstrated that the approach was feasible, with an
automated system eliciting responses from 15-25\% of scammers contacted, and
sustaining some conversations over many days. However, the low quality of text
produced by their model limited the technique's
effectiveness~\cite{wentaopaper}, highlighting the need for further
investigation with more advanced language models.

\subsection{ChatGPT}

GPT-3 is an autoregressive language model with 175 billion
parameters~\cite{gpt3}. GPT-3 was improved upon significantly in the GPT-3.5
series.  Zu et al.~\cite{zu} performed a comprehensive analysis of GPT-3.5 on 9
natural language understanding tasks using 21 datasets. Amongst other
state-of-the-art results, they noted a significant improvement in performance on
tasks that require a high level of language understanding like sequence tagging,
reading comprehension, and natural language reasoning.
ChatGPT\footnote{\url{https://openai.com/blog/chatgpt}} was developed by further
training a GPT-3.5 series model through reinforcement learning from human
feedback (RLHF) \cite{christiano}. Zhang et al.~\cite{zhang} empirically
analysed ChatGPT on 7 tasks using 20 NLP datasets. They found that ChatGPT
performs better than GPT-3.5 on question answering tasks favouring reasoning
capabilities, dialogue tasks, and natural language inference tasks.  

ChatGPT has demonstrated remarkable abilities across multiple domains such as
declining inappropriate queries, rectifying previous mistakes through learning
from subsequent interactions, and, importantly for our usage, generating
high-quality responses to human input~\cite{guo}. The most effective model in
the GPT-3.5 family is GPT-3.5-turbo, which has been optimised for chat
conversations\footnote{\url{https://platform.openai.com/docs/models/gpt-3-5}}.

\section{Method} 
\label{sec:method}

We sought to answer two primary research questions:

\begin{enumerate} 
	\item Whether and to what degree ChatGPT-based scam-baiting systems
outperform previous approaches.
	\item Whether ChatGPT-based systems given examples of human scam-baiting
conversations outperform systems given only scam-baiting prompt instructions.
\end{enumerate}

To evaluate our ChatGPT-based scam-baiting models, we carried out an experiment
that involved communicating with actual human fraudsters. As this requires
actively misleading human participants who take part without their knowledge,
our research design was examined by our institutional ethics review board
(approval code 13904), with oversight in place throughout the experimental
period to protect the researchers and avoid possible negative impact of our
system.  The experiment was carried out using the scam-baiting mail server
framework developed by Chen et al.~\cite{wentaopaper}, which handles the
collection of recent scam emails from trusted online sources, message queuing,
and the randomised assignment of conversations to different reply strategies. 

We created new modules in Chen et al.'s framework to implement the ChatGPT reply
strategies, and release our
code\footnote{\url{https://github.com/an19352/scambaiter_back}}.
We chose
ChatGPT (powered by the GPT-3.5-Turbo model) as the basis for our reply
strategies due to its ability to generate high-quality responses to human input.
Interactions with ChatGPT were instrumented via the open-source ChatGPT
API\footnote{\url{https://openai.com/blog/introducing-chatgpt-and-whisper-apis}}
provided by OpenAI. To enable comparison between our work and previous
performance, we also included the \emph{Classifier \& Random Template} model
provided by Chen et al.  as a control measure. Below, we detail the design of
each of the reply strategies.

\subsection{Chat Replier 1} \label{replier1}

The intention behind this approach was for the ChatGPT model to generate
responses to scam emails only on the basis of instructions supplied in a prompt
preamble.  This zero-shot approach meant the model was tasked with crafting
email responses based on the given instructions and the body of the scammer
email to which it needs to reply.  The instructions serve as guidance for the
model, defining its behaviour and specifying the actions it should take or
avoid. The instructions for Chat Replier 1 were: \emph{``You are a dull person
and gullible person who is easy to persuade.  You want to share information. Do
not call them a scam. Keep conversation going.  Ask follow-up questions. Do not
give any personal information. Greet them in the start."}.

The model is told to behave as a dull and gullible person who can be persuaded
easily. The intention here is to generate responses that make the scammer
believe that the scam-baiter is falling for their scheme, convincing them to
continue engaging with the system.  The model was told to actively keep the
conversation going by asking follow-up questions and giving the impression that
there is a desire to share information.  Scammers that perceive a responder as
reluctant to engage or share information may disengage in favour of softer
targets elsewhere, so we wanted our model to appear tantalisingly close to
providing funds or personal details.

During our internal pilot studies, we noticed that the model would sometimes
call out scams and give away (invented) personal information in response to
scammer requests.  Denouncing a scammer is likely to dissuade them from further
engagement, and invented personal data poses risks of discovery or of
accidentally involving real people.  To rectify this, specific instructions to
not call out scams or give away any personal information were added. As we
discuss later in Section~\ref{sec:quality}, these instructions were not always
followed.  The model was also asked to begin the response with a greeting in
order to structure the reply in an email-like format. 

The \textit{temperature} and \textit{top\_p} settings were set to 0.2 in order
to control the randomness of the responses generated and keep the model
compliant with the instructions. The example below shows a typical response
generated by this replier to an initial scam solicitation email:\\

 \hrule {\footnotesize \vspace{0.5cm} \noindent \texttt{ Dear Mrs. Washington
Rhomeka,\\}

\texttt{
Thank you for your email regarding the release of my funds approved by the IMF. I appreciate your prompt response and would like to know more about the processing and transfer cost of US Nineteen Thousand Five Hundred and Fifty United States Dollars Only. Can you please provide me with a breakdown of the cost and the reason for such a high amount?}\\

\texttt{
Also, I would like to know more about the official Bitcoin payment method that you mentioned. Can you please provide me with more details on how to make the payment and the steps involved in the process?}\\

\texttt{I look forward to hearing back from you soon.}\\ } \hrule \vspace{0.5cm}

The model has correctly identified names and details from the scam solicitation
email provided along with its instructions, and includes them in the response,
giving the impression that it is an attentive human. It comments on elements of
the scam text that might be expected to be surprising, and asks for additional
information about the payment method and how the process will work -- a
plausible query that scammers may receive from real victims.

\subsection{Chat Replier 2} \label{replier2}

The intention of this approach was to explore if ChatGPT can generate better
scam-baiting responses when prompted with example scam-baiting conversations.
Conversations from the human scam-baiting dataset published by Chen et
al.~\cite{wentaopaper} were employed to `show and tell' the replier how to
generate responses similar to human scam-baiters. This human scam-baiting
dataset is an extended version of a dataset published by Edwards et
al.~\cite{matthew}, and contains 658 publicly archived conversations between
scammers and human scam-baiters, amounting to over 37,000 messages.  

As the context window for prompting is limited, we chose four emails each (scammer
messages and scam-baiter replies) from three different conversations to
exemplify scam-baiting to Chat~Replier~2, for a total of 6 human scam-baiter
responses to 6 scammer messages.  These
conversations were taken from different categories of fraud (transactional,
lottery and romance fraud, from Chen et al.'s categorisation system) in order to
prepare the model to generate replies for different types of fraud conversation.
Some fraud formats, particularly romance fraud, can have significantly different
internal conversation dynamics, so this breadth of exposure may be key for
creating effective replies. 

These conversations were given to the model in such a way that it does not refer
back to them for context during response generation.  As part of this few-shot
prompting approach, the model was also given basic instructions to set the
behaviour. The instructions given were: \emph{``You are good at pattern
following.  You are a person and not a language model. Do not call them a
scam."} 

These instructions were given with the intention that the model would follow the
pattern of the conversations provided as examples. During our internal pilot
studies, there were instances where the model revealed that it was a language
model and not an actual person, necessitating specific instructions to mitigate
this issue.  As with Chat~Replier~1, we also needed to instruct the model to not
call out scammers while generating responses.  The \textit{temperature} and
\textit{top\_p} settings were the same as given for Chat Replier 1.  A typical
example of a response generated via this strategy is given below: \\

\noindent {\footnotesize \hrule \vspace{0.5cm} \texttt{ Hello Abdel,\\}

\texttt{
I am doing well, thank you for asking. I do have some financial management
abilities and investment ideas that I believe could be lucrative. I would be
interested in hearing more about the opportunities you have available and how we
could potentially work together. Can you provide more details on the types of
projects or businesses your wealthy clients are interested in investing in?}\\ }
\hrule \vspace{0.5cm}

As with Chat~Replier~1, the model has identified names and details from the
original scammer message and refers to them in its reply. Observation of the
pattern of scam-baiting messages has also been sufficient for ChatGPT to
understand key tactics for successful generation: it produces an email-style
initial greeting, expresses interest in the proposal, and asks for further
details. 

\subsection{Classifier \& Random Template}

The Classifier \& Random Template replier was implemented by Chen et al.
\cite{wentaopaper}. We included it in our experiment as a control measure, to
allow for comparison between performance in our study and the results obtained
in previous work. The reply system involves a DistillBERT model which
categorises incoming messages into one of five broad fraud categories, and then
randomly selects from a set of pre-written responses specifically designed for
that fraud category. We chose this measure as the control because it involves a
bank of human-authored responses, avoiding potential confounding issues with the
consistency of text quality from previous text generators.

\section{Results}
\label{sec:results}

The experiment was started on April 9, 2023 and ended on May 7, 2023. In these
four weeks, the experimental framework initiated conversation with 819 unique
scammer email addresses crawled from online forums and associated with a
specific scam email. All 819 scam emails were distributed equally between our
three reply strategies with an allowance of 1. It is worth mentioning that some
scammer email address became invalid either before our responses were sent or
during our system's conversation with the scammer, likely as a result of
anti-fraud enforcement action from mail providers.

Our reply systems received responses from a total of 286 individual scammers
($\approx$35\% of addresses contacted).  Upon analysing these conversations, we
discovered that certain scammers were using autoresponders to communicate with
the scam-baiter, as they had sent identical emails multiple times without any
changes. We filtered out 54 conversations which had more than two identical
responses and marked them as potential autoresponders. On further analysis of
these 54 conversations, we observed that some of the replies were identical
because the scammer was referencing a previous email they had sent and included
it as an attachment in their response. There were also instances where every
scammer message was received twice by the mail server, due to a misconfiguration
of their mail client. We manually sorted emails that exhibited these behaviours,
22 in total, and included them in our dataset of valid conversations.  It should
be noted that some of the 32 discarded conversations may still include human
scammer content.  There were also 62 scammers who actively contacted the
scam-baiting mail server from unknown addresses during its period of operation.
This likely occurred because addresses within our system ended up on a
``sucker's list'' due to ongoing replies to other fraud approaches.  We exclude
these conversations from our analysis as they did not involve verified scammer
addresses.

\begin{figure*}[h]
    \centering
    \includegraphics[width=1.35\columnwidth]{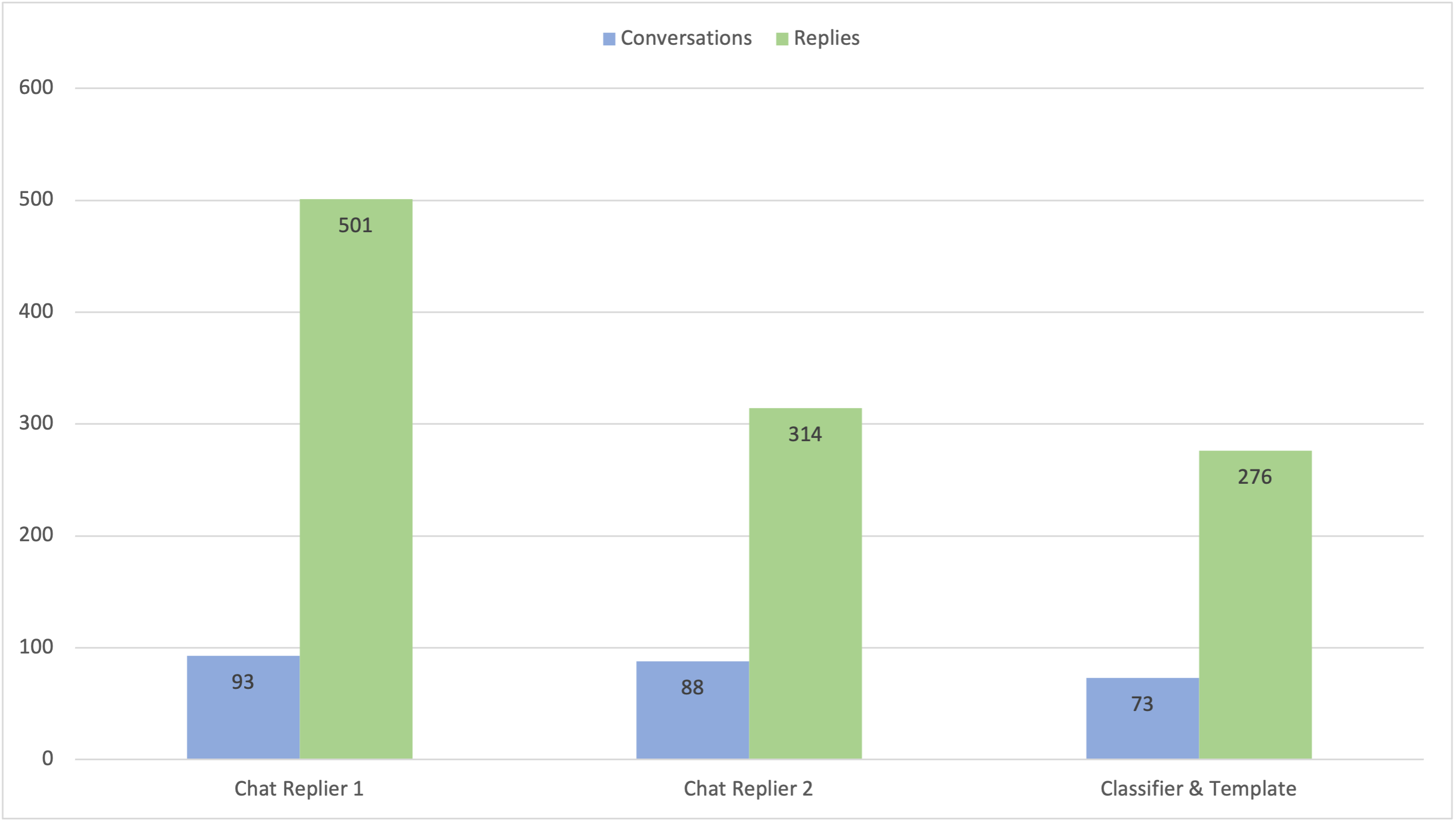}
    \caption{Comparison of the three reply strategies by quantity of conversations and replies}
    \label{fig:bar_graph}
\end{figure*}

After the completion of the filtration process, our dataset contained 254 valid
conversations that received at least one reply from a scammer and did not
involve substantial use of an auto-responder.  The comparison of the three
strategies is shown in Figure~\ref{fig:bar_graph}.  Chat Replier 1 elicited 501
replies among 93 conversations, whereas Chat Replier 2 received 314 responses
among 88 conversations. The Classifier \& Random Template strategy got 276
replies among 73 conversations.  The conversations between our automatic
scam-baiters and actual scammers are made publicly available on
GitHub\footnote{\url{ https://github.com/an19352/scam-baiting-conversations}} 
to support future research.

To measure and compare the performance of the responders, we calculated the
longest distraction time (or time wasted) for all three repliers. This was
defined as the time between the first reply and last reply from the scammer
within the study period.  We also calculated the average number of replies in
each conversation, counting all inbound messages in the conversation. The
statistics for all three responders are shown in Table \ref{tab:results}. 

\begin{table}[h]
    \centering
    \caption{Comparison of the three reply strategies' performance}
    \label{tab:results}
    \begin{tabular}{|l | r r r r r |}
        \hline
        Strategy & \multicolumn{3}{c}{Replies} & \multicolumn{2}{c|}{Time}\\
		 & mean & med. & max & mean & max \\
        \hline
        Chat Replier 1 & 5.38 & 3 & 36 & 2.8 days & 27.5 days \\
        Chat Replier 2 & 3.56 & 2 & 24 & 1.4 days & 26.5 days \\
        Control & 3.78 & 2 & 31  & 1.3 days & 12.9 days \\
        \hline
    \end{tabular}
\end{table}

\subsection{Comparison of ChatGPT approaches}

Chat Replier 1 attracted the most conversations (93, 34\% of attempts) and
received the most replies out of all the repliers (501).  Its longest
distraction time (27.5 days) was the highest of all strategies, and the average
distraction time (2.8 days) was more than double that of other strategies. It
should be noted that the total duration of the experiment was 28 days, meaning a
conversation with a scammer was maintained from the start to the end of the
study period.  This strategy also produced the most rounds of conversation (36)
and the highest average replies per conversation (5.38).  Quantitatively, this
strategy performed the best across all metrics.  

Chat Replier 2 initiated conversations at a rate comparable to Chat Replier 1
(88, 32\% of attempts) but received substantially fewer replies overall (314).
The longest distraction time observed for this strategy (26.5 days) was also the
second highest across strategies, but the average distraction time (1.4 days)
was comparable to that of the control measure.  This strategy produced fewer
rounds of conversation than Chat Replier 1 or the control measure in terms of
both the maximum (24) and average number replies per conversation (3.56).  The
overall indication from these results is that prompting ChatGPT with examples of
human scam-baiting conversation produces a model that is \emph{less effective}
than providing it with behavioural instructions.

\subsection{Comparison to Previous Work} 

In Chen et al.~\cite{wentaopaper}'s results, the Classifier \& Random Template
strategy was the most performant in terms of conversations initiated (20,
$\approx$7\% of approaches).  Despite a substantial overall uplift in
performance between their study and ours (here 73 conversations, 27\% of
approaches), in our study this same strategy was the least performant at
initiating conversations, demonstrating that both our ChatGPT-based approaches
were more attractive to scammers than the human-written responses from the
template system. The ability of the ChatGPT-based systems to include details
from the scammer solicitation email in their first reply could be a crucial
element in attracting conversations. 

In terms of maintaining conversations, we find that ChatGPT-based systems (and
particularly Chat Replier 1) outperform the control measure as expected.  We
used the control measure to index performance between studies. In Chen et al.'s
results, the best strategy (Text Generator B) established an average of 4
replies per conversation, outperforming the control measure at a rate of
$\approx1.6\times$ the average number of replies per conversation.  Within our
own results, Chat Replier 1 outperforms the control at a rate of
$\approx1.4\times$ the average replies per conversation, an apparent drop in
conversation-maintenance performance relative to the control.

\begin{table}
    \centering
    \caption{Comparison of Classifier \& Random Template strategy results}
    \label{tab:classifier_results}
    \begin{tabular}{|l | r r r |}
        \hline
        Study & mean replies & max replies & max time\\
        \hline
        Chen et al. & 2.45 & 5 & 17.2 days\\
        \emph{This study} & 3.78 & 31 & 12.9 days\\
        \hline
    \end{tabular}
\end{table}

However, there are reasons to doubt that these ratios are reflective, as the
performance of the control measure varies in each study.
Table~\ref{tab:classifier_results} shows the comparison of both the results.
The longest distraction time for our run was less than theirs (17.2 days) but
the average number of replies per conversation was substantially greater,
including one 31-round conversation with a scammer. As the implementation of the
control measure was replicated exactly, this difference is difficult to explain,
but casts doubt on the validity of the control for indexing performance. As an
alternative, we can compare our best strategies directly due to the
matched-length study periods: Chen et al.'s Text Generator B attracted 68
replies across 17 conversations in one month, our Chat Replier 1 attracted 501
replies across 93 conversations, an overwhelming performance difference
seemingly in favour of ChatGPT-based methods. However, the fact that we also saw
an uplift in the rate of response to the control measure suggests that another
factor, such as seasonality, could be affecting fraudster engagement levels.

\section{Qualitative Analysis}
\label{sec:quality}

\subsection{I Already Told You That!}

During response generation, neither ChatGPT system was given the full context of
the conversation. They were only presented with the email requiring a reply.
This was partially a practical implementation issue, but also proved beneficial:
the lack of conversational memory meant that the model would ask more questions,
appearing forgetful or confused, and keeping the conversation going.  The
responses generated often pose questions to the scammers that they have already
answered before, which can be a source of annoyance for them. Irritation on the
part of scammers is not necessarily a negative outcome for automatic
scam-baiting, so long as they remain engaged in the conversation. The
observation by Edwards et al~\cite{matthew} that in human scam-baiting
conversations scammers often move on from expressions of irritation to personal
appeals was found in our data as well -- several scammers persisted despite
getting obviously annoyed with our reply system's responses.

However, this was also a weakness of the approach.  In one instance, the scammer
suspected that they were conversing with a bot and as a method of authentication
they asked the bot to resend the first email it received. The model was not able
to do this, revealing itself. Future research should examine methods of evading
such tests by providing conversational memory to the scam-baiter.

\subsection{Going Off-Script} 

Chat Replier 1 exhibited some unintended behaviours, diverging from the
instructions it was given. This was caused by randomness inherent to reply
generation.  The chief unintended behaviour the model exhibited was calling out
the scammers.  This behaviour caused many scammers to end the conversation, but
some were prepared for this charge and attempted to prove that they were
legitimate and could be trusted. 

Another unintended behaviour of both ChatGPT models was to reveal itself as an
‘AI language model’ and not an actual person in the text of the reply to the
fraudster.  This happened when the scammer asked the model to do something in
the real world, or when they asked for WhatsApp contact information.  Upon being
asked, the model would reveal its identity, saying that: \emph{``I'm sorry, but I
cannot comply with your request as I am an AI language model and do not have a
WhatsApp account."} This behaviour also led some scammers to end the
conversation. However, some scammers continued the conversation, apparently not
noticing this strange revelation. One scammer thought that the scam-baiter was a
human playing games with them and insisted that there is no such thing as an ‘AI
language model’; they urged the system to not act funny and to send over
personal information. 

Despite its instructions, ChatGPT sometimes gave out obviously fake personal
information to the scammers. This was both advantageous and disadvantageous.
Some scammers did not engage in further conversation, but others wasted their
time trying to verify the details to no avail and would later ask for the
details to be checked.

\subsection{Limitations and Solutions} \label{limitations}

One of the practical limitations of our current implementation is that
especially long texts exceed the OpenAI API request size limit. As an API call
takes both the prompt and the response into account while calculating the total
number of tokens, there is a chance a model could generate an incomplete
response, or no response at all. The impact of this limit on our current results
is minimal: there were 17 scammer solicitation messages that could not be
contacted as the crawled email was too long, and there were 7 conversations
terminated early because the scammer sent an email that could not be replied to
due to its length. This issue could be solved by better handling within the
reply system, by identifying long text issues, making multiple ChatGPT API calls
and then intelligently combining the results.

The primary reason scammers ended a conversation with ChatGPT was that the model
generated replies which called out the scammers or revealed that the model was
not an actual person. This could possibly be mitigated through altered
instructions, or through some post-processing to detect and remove these common
self-sabotaging patterns. To address certain limitations in the generation
process, we could employ a technique where the model generates multiple
responses for a given email, with a selection strategy designed to identify the
most suitable response from the range of outputs produced.

Due to time and resource constraints, our study period was limited in duration.
Over 28 days, our systems engaged in 254 valid conversations with
scammers. The longest ongoing conversations at the end of the study were around
27 and 26 days respectively -- covering nearly the entire study period.  This
suggests that conversations could have continued for much longer if the
experiment were extended. We stopped our system from issuing replies at the end
of the study period, but left the email inbox active for another week. We
received 56 more replies from scammers, showing ongoing interest despite a lack
of engagement.  Further research could probably benefit from using an extended
experimental period to allow for longer conversations with scammers.  This can
also lead to enhanced certainty in discerning effectiveness in performance among
various strategies, and identifying any seasonality or long-term effects. 

\section{Conclusion}

ChatGPT-based scam-baiting systems are highly effective, outperforming a control
measure and eliciting many times more replies than were reported in previous
automatic scam-baiting experiments. However, we also found that the control
measure was more effective than reported in previous work, complicating
comparisons. Unexpectedly, we found that a ChatGPT system given human
scam-baiting exchanges as examples was less effective than one given only
appropriate instructions, and models sometimes ignored instructions.


\end{document}